\documentclass[11pt]{amsart}
\usepackage{geometry}                
\usepackage[parfill]{parskip}    
\usepackage{graphicx}
\usepackage{amssymb}
\usepackage{epstopdf}
\usepackage{color}
\newcommand{\R}{\mathbb{R}}
\usepackage{float}
\title[Statistical Methods in Topological Data Analysis]{Statistical Methods in Topological Data Analysis for Complex, High-Dimensional Data}
\author{Patrick S. Medina \& R.W.\ Doerge}

\begin{document}

\maketitle

\begin{abstract}
	The utilization of statistical methods an their applications within the new field of study known as Topological Data Analysis has has tremendous potential for broadening our exploration and understanding of complex, high-dimensional data spaces.   This paper provides an introductory overview of the mathematical underpinnings of Topological Data Analysis, the workflow to convert samples of data to topological summary statistics, and some of the statistical methods developed for performing inference on these topological summary statistics.  The intention of this non-technical overview is to motivate statisticians who are interested in learning more about the subject.
\end{abstract}

\section{Introduction}

\begin{figure}[b] 
	\includegraphics[scale=0.5]{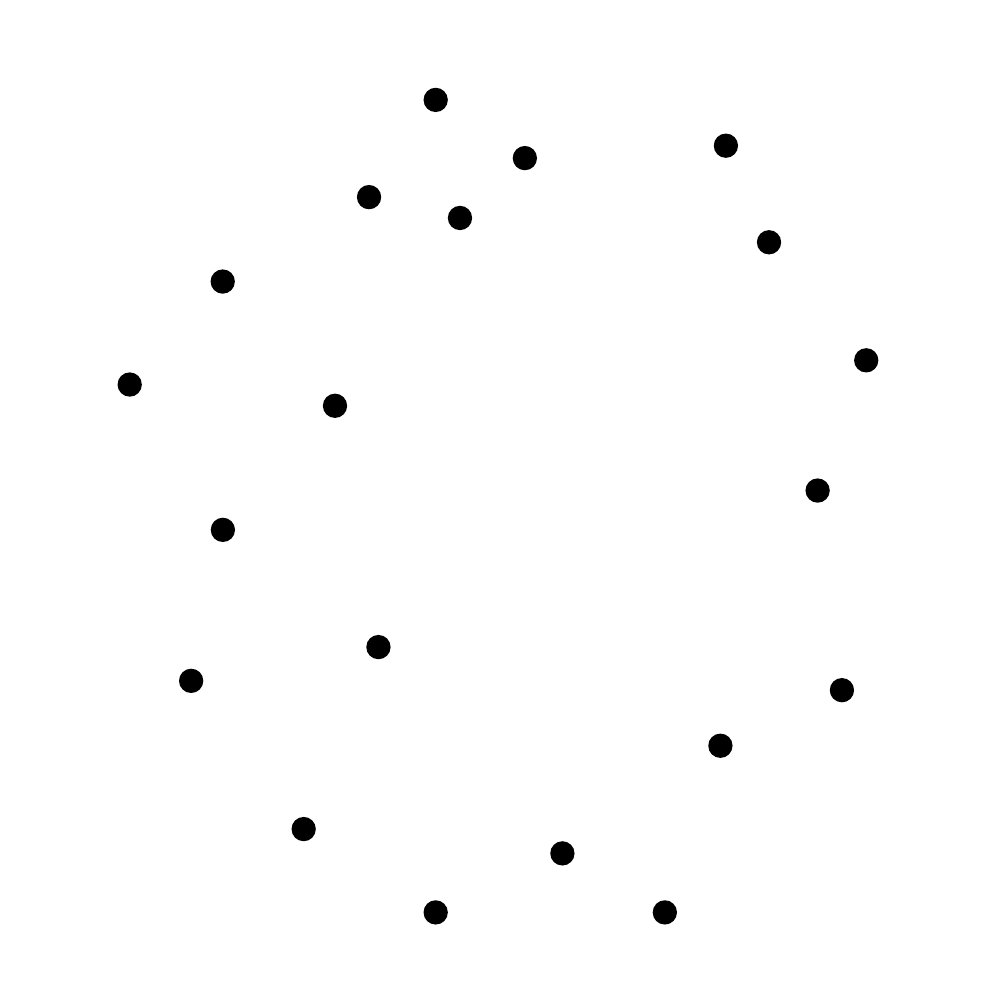}
	\caption{A random sample of 20 points sampled from some distribution $\mathcal{F}$ with unknown support $\mathcal{X} \subseteq \R^n$.  }  \label{sample}
\end{figure}

Suppose one is asked to analyze the sample data in Figure \ref{sample}.  What could be said?  Obviously, it is two dimensional, and it appears to be circular.  However, it may be that these data were sampled from a distribution whose support is in the shape of a circular coil.  Toward this end, how could a sample of data points be used to study the overall shape of the support of the distribution from which they were sampled?  Furthermore, is it possible to learn this if these data are high-dimensional?

Topological Data Analysis (TDA) has emerged as a branch of computational topology that enables researchers to study the shape properties of a mathematical space based on a representative sample taken from that space.  The information is then used to learn how the original mathematical space is organized.  TDA uses ideas from algebraic topology to quantify distinct shape characteristics of mathematical spaces.  These concepts are general enough that they extend to data that are high-dimensional and very complex; i.e. they reside in spaces where traditional linear statistical methods or manifold learning techniques may fail to adequately capture properties of the underlying space.  

In the context of a statistical problem, it is assumed that a sample of data $\mathcal{P}$ is drawn randomly from some distribution $\mathcal{F}$ whose support, $\mathcal{X} \subseteq \R^n$, is unknown.  Based on this setting, tour motivation is to provide statisticians with a very friendly introduction to the mathematical underpinnings of Topological Data Analysis.  Specifically, we provide an overview of the process by which data, $\mathcal{P}$, are converted to different topological summary statistics. The statistical methods developed for inference on a sample of topological summary statistics are introduced.  Finally, the application of some of these statistical methods in analyzing differences between two structures of the maltose binding protein are examined.

\begin{figure}[t] 
	\includegraphics[scale = 0.47]{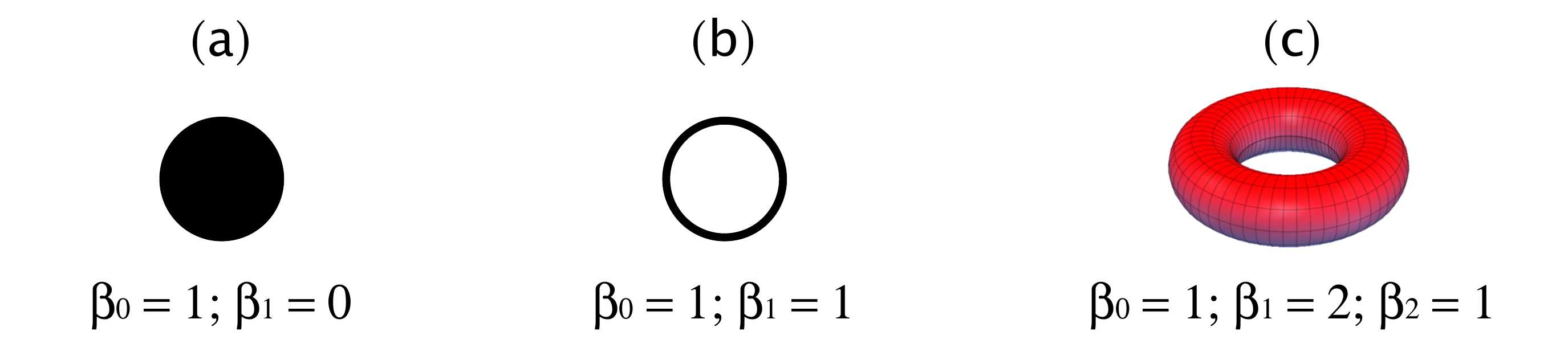}
	\caption{Three different mathematical spaces and their Betti numbers.  Betti numbers quantify the distinct number of shape features that appear in a mathematical space.  Specifically, connected components are $H_0$, loops by $H_1$, voids by $H_2$, and so on for higher dimensional analogues.  \textbf{(a)} The image of a disc embedded in $\R^2$.  If any two points are chosen in the disc then a line may be drawn connecting the two points.  Hence, there is one connected component and the Betti number for $H_0$, $\beta_0$ is one.  Further, since there are no holes in the disc then the Betti number for $H_1$, $\beta_1$ is one.  \textbf{(b)}  The image of a circle embedded in $\R^2$.  If any two points are chosen along the circle then they may be connected by an arc between them.  Also, there is a large hole inside of the circle.  Hence, $\beta_0$ and $\beta_1$ are one.  \textbf{(c)}  The image of a Torus embedded in $\R^3$.  If any two points are chosen along the exterior, then they may be connected by drawing a path between them, hence $\beta_0$ is one.  A torus has two holes: the first is the large one in the center, and the second is seen by cutting the torus in half.  Hence, $\beta_1$ is two.  Finally, the inside of the torus is hollow, which means it has one void and the Betti number for $H_2$, $\beta_2$ is one.}  \label{betti-num}
\end{figure}

\section{Overview of Topological Data Analysis}

TDA combines methods from the fields of algebraic topology and computational geometry for the purpose of studying key shape features of a mathematical space from which a sample of data may have been drawn.  Two key tools for achieving this are homology and simplicial complexes.


\subsection{Homology}  Homology is a subject in algebraic topology that provides tools for computing the number of distinct shape features within a mathematical space.  The shape features of interest are connected components - notated by $H_0$, loops - notated by $H_1$, voids - notated by $H_2$, and their higher dimensional analogues.  For each of these shape features a Betti number quantifies the distinct number of shape features that appear in a mathematical space \cite{carlsson_topology_2009}.  An example of different mathematical spaces and their Betti numbers is illustrated in Figure \ref{betti-num}.  Specific details on homology can be found in Munkres \cite{munkres_elements_1984}.

Betti numbers are typically used to understand the shape characteristics of a mathematical space.  However, since data are a collection of discrete points, using the data directly as a representation of the underlying mathematical space will not, in general, capture interesting features that may exist.  Hence, in order to learn about the shape features of the mathematical space, a geometric representation, known as a simplicial complex, of the space needs to be constructed from the data.


\subsection{Simplicial Complexes}  
\begin{figure}[htb]
	\includegraphics[scale=.75]{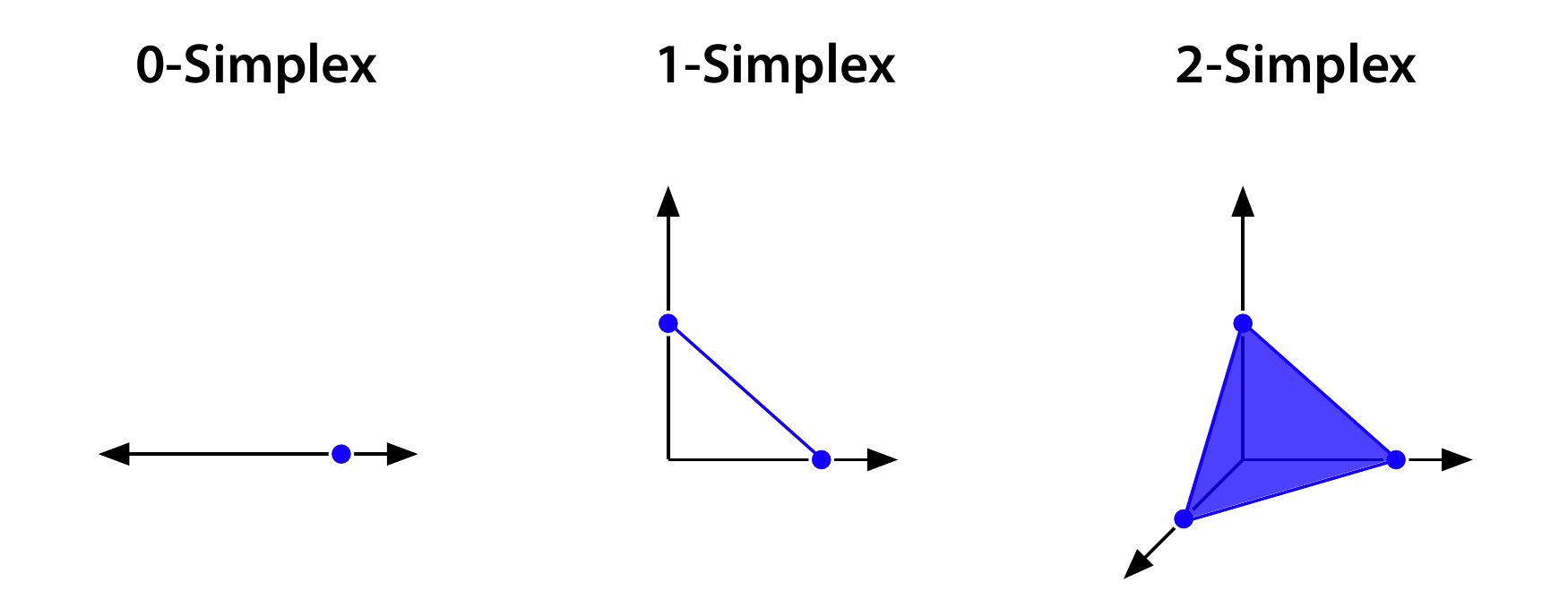}
	\caption{Examples of the first three simplices.  The 0-simplex embedded in $\R$ is a point.  The 1-simplex embedded in $\R^2$ is a line segement.  The 2-simplex embedded in $\R^3$ is a triangle.} \label{simplices}
\end{figure}

The main tool for constructing a geometric representation of a mathematical space from data is the simplicial complex.  These are topological structures constructed by attaching simplices along their faces.  A simplex is a general term for triangles in a Euclidean space.  The 0-simplex, or vertex, is a point.  The 1-simplex is a line segment. The 2-simplex is a triangle.  Higher dimensional equivalents of these objects follow directly.  Examples of the first three simplices are illustrated in Figure \ref{simplices}. Computational methods have been developed to construct a simplicial complex using a sample of data as the vertices.  The details of simplicial complexes can be found in Munkres \cite{munkres_elements_1984}.

\subsubsection{Computational methods for constructing simplicial complexes} \label{VRComplex}    One of the most common approaches for constructing a simplicial complex is the Vietoris-Rips complex \cite{zomorodian_fast_2010}.  It is constructed by examining all pairwise distances between points, $\mathcal{P} = \{p_1,p_2, \dots, p_n \}$.  Since the data are from a subset $\mathcal{X} \subset \mathbb{R}^n$, numerous notions of distance can be used.  Two common metrics are the $p$-distance,
\begin{equation*}
	d_p(x,y) = \Big( \sum_{i = 1}^{n} |x_i - y_i|^p \Big)^{\frac{1}{p}} \, ,
\end{equation*} 
and the maximum distance,
\begin{equation*}
	d_\infty(x,y) = \max \Big \{|x_1 - y_1|, |x_2 - y_2|, \dots, |x_n - y_n| \Big\} \, .  
\end{equation*} 
When $p = 2$, the $p$-distance is the Euclidean distance that is used in many applications, such as regression.  While these metrics are commonly used in the construction of the complex, other metrics may certainly be considered.

The Vietoris-Rips complex is most easily understood by considering a subcollection of points of $\tau =  \{p_{i_1},\dots,p_{i_m}\} \subseteq \mathcal{P}$.  $\tau$ is a simplex in the simplicial complex if the points are all relatively close to each other.  Specifically, the concept of closeness is conceived via a fixed scale parameter $\epsilon > 0$, and $\tau$ is a simplex in the simplicial complex if $d(p_{i_j},p_{i_k}) < \epsilon$ for all $i_j$ and $i_k$.  Once this complex is constructed, it is possible to use the tools developed from homology to compute the number of distinct shape features present.

\subsubsection{Choosing an appropriate scaling parameter}  
\begin{figure}[htb!] 
	\includegraphics[scale = 0.5]{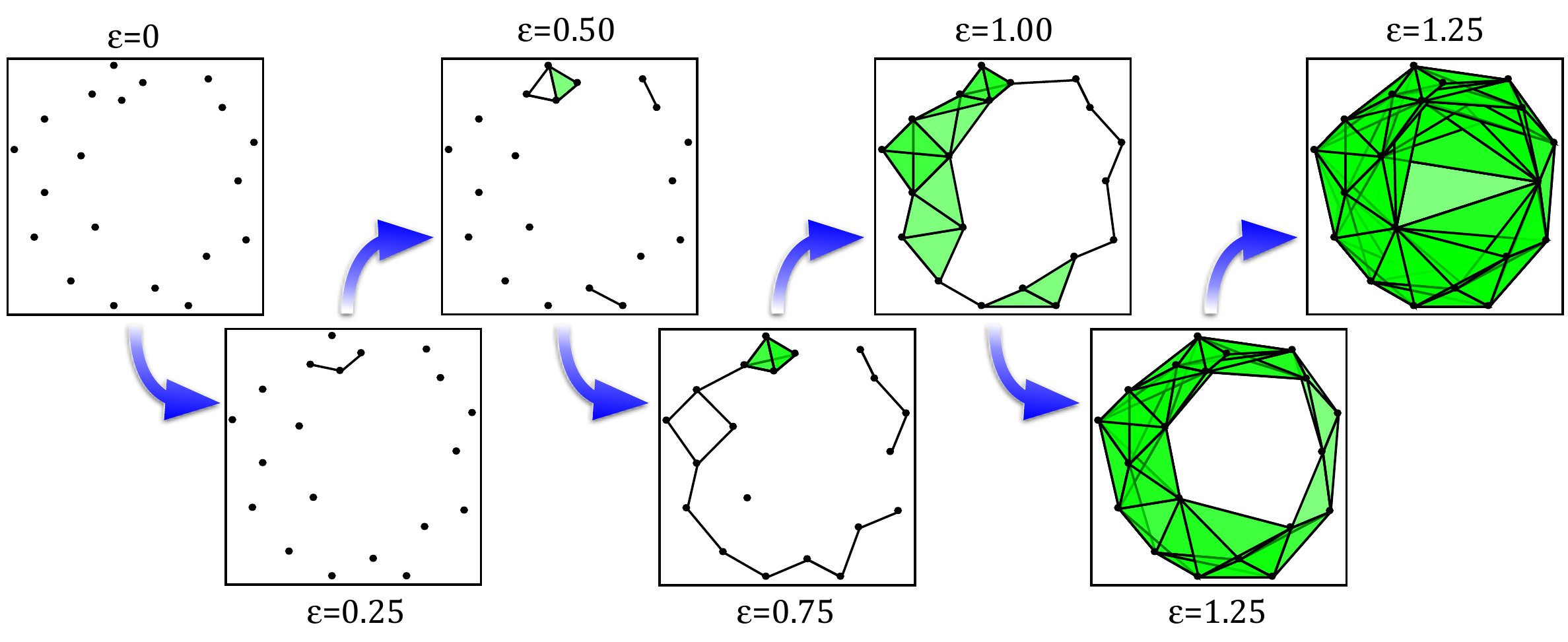}
	\caption{The evolution of the simplicial complex at different values of the scaling parameter.  As the scaling parameter increases, more simplices are added to the simplicial complex.} \label{simplexfilt}
\end{figure}

For a fixed value of $\epsilon$, the Vietoris-Rips complex is an approximation of the underlying structure.  As $\epsilon$ increases, more simplices are added to the simplicial complex, until it contains all possible simplices that can be generated from $\mathcal{P}$. Because of the dependency on the different values of the scale parameter, information about the shape of the underlying mathematical space varies, see Figure \ref{simplexfilt}.  

Instead of using a specific scale parameter $\epsilon$, Topological Data Analysis uses a range of scale parameters to dynamically keep track of when distinct shape features appear and disappear from the simplicial complex.  
The concept is similar to the idea of hierarchical clustering, which tracks cluster membership over a specified scale parameter, and is known in TDA as ``persistent homology.''

\subsection{Persistent Homology}  As mentioned, persistent homology \cite{edelsbrunner_topological_2002} tracks the appearance and disappearance of distinct shape features in a simplicial complex via a changing scale parameter.  It allows users to understand the number of features that appear in their data, but also, it evaluates how long these features exist.  The details of persistent homology may be found in Edelsbrunner et al. \cite{edelsbrunner_topological_2002} and Edelsbrunner and Harer \cite{edelsbrunner_computational_2010}.  Reviews of the Topological Data Analysis workflow may be found in Carlsson \cite{carlsson_topology_2009}, Ghrist \cite{ghrist_barcodes:_2008}, and Nanda and Sazadanovi\'{c} \cite{nanda_simplicial_2014}.


\section{Topological Summary Statistics}

Topological summary statistics allow the quantification and visualization of persistent homology.  The main summary statistics used in Topological Data Analysis are the persistence diagram, barcode, and the persistence landscape.  Persistence diagrams and barcodes are closely related and, as such, the focus here will be on the persistence diagram and persistence landscape.


\subsection{The Persistence Diagram}  Persistent homology tracks the existence of distinct shape features over a range of values for a scaling parameter.  If a shape feature appears at a parameter value of $\epsilon_a$, and disappears at a parameter value of $\epsilon_b$, then then the persistence diagram retains this information through the point $(\epsilon_a,\epsilon_b)$ in $\R^2$.  The length of the shape feature's existence can be measured by the vertical distance between this point and the diagonal ($y = x$).  Hence, a persistence diagram is a multiset (a set that allows the number of elements to be repeated) of points in $\mathbb{R}^2$, and the diagonal, where the diagonal has infinite multiplicity.  Figure \ref{summary-stats} illustrates a persistence diagram and explains its interpretation. Details and methods for computing persistence diagrams are covered in Edelsbrunner and Harer \cite{edelsbrunner_computational_2010}.


\subsubsection{Mathematical properties of the persistence diagram} \label{math-pd}  In order to make the space of persistence diagrams amenable to statistical inference, the definition of a persistence diagram is limited to a finite multiset of points in $\R^2$, and the diagonal $(y = x)$, where each point on the diagonal has infinite multiplicity \cite{mileyko_probability_2011}.

The set of persistence diagrams that satisfy this definition, $\mathcal{D}$, equipped with the Wasserstein metric is considered a metric space.  The $p^{th}$ Wasserstein distance between two persistence diagrams $d_1, d_2 \in \mathcal{D}$ is defined as
\begin{equation*}
	W_{p}(d_1,d_2) := \Big( \inf_{\gamma} \sum_{x \in d_1} ||x - \gamma(x)||_\infty^p \Big)^{\frac{1}{p}} \, ,
\end{equation*}
where $\gamma$ varies across all one-to-one and onto functions from $d_1$ to $d_2$.  When $\mathcal{D}$ is restricted to the set of persistence diagrams whose $p^{th}$ Wasserstein distance from itself to the diagonal is finite, Mileyko et al. \cite{mileyko_probability_2011} show that the space of previous persistence diagrams is a Polish space \cite{dudley_real_1989}.  This result gives rise to the statistical construct of a mean and variance for persistence diagrams (see Section \ref{mean-diagram}).


\subsubsection{Comparison to the barcode}  
\begin{figure}[H]
	\includegraphics[scale=0.60]{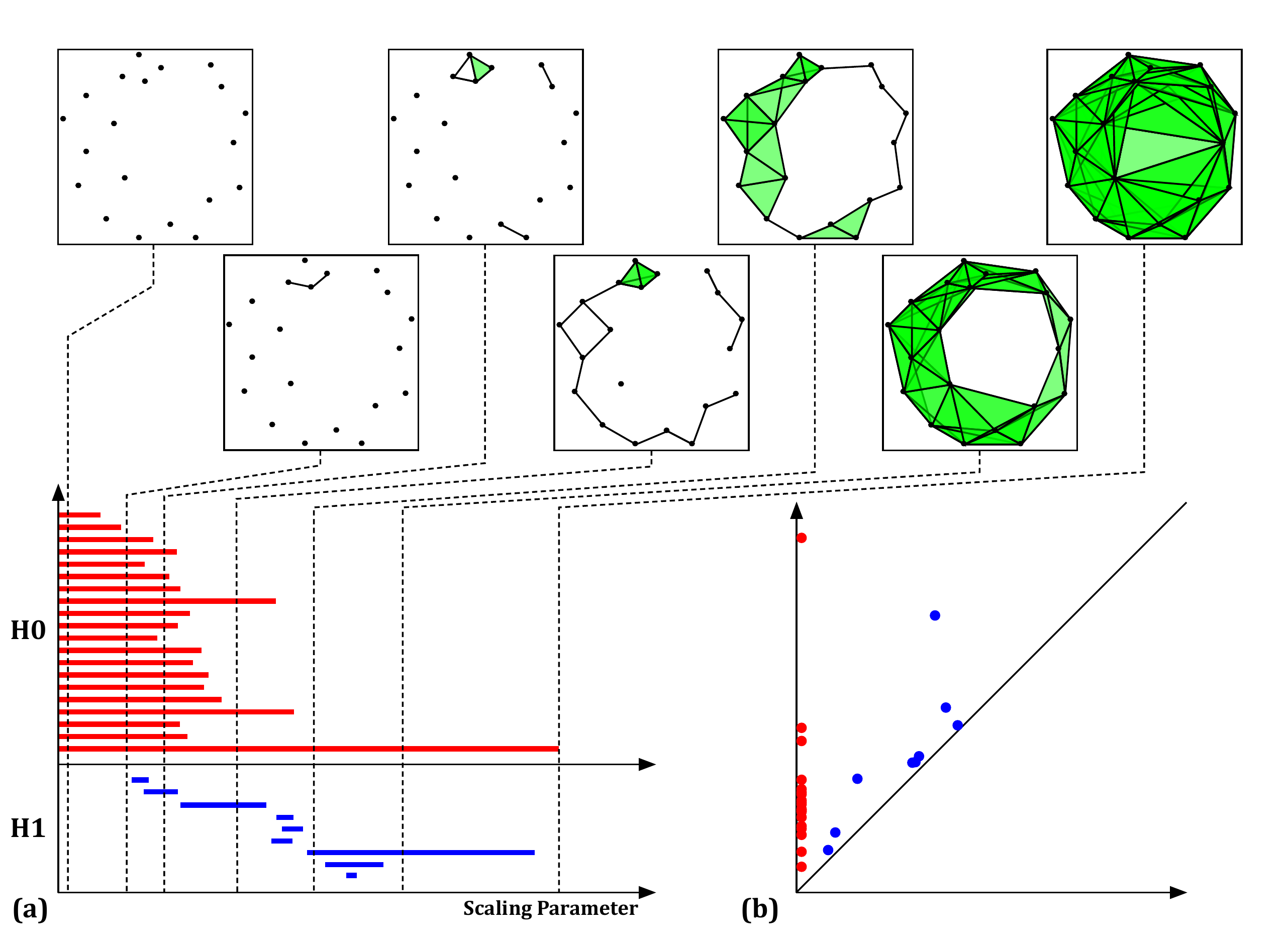}
	\caption{The top of the plots represent the simplicial complex at different values of the scaling parameter.  \textbf{(a)} Illustration of the barcode summary statistic.  The red and blue bars show the persistent homology of the connected components, $H_0$, and loops, $H_1$, respectively.  The $x$-axis is the scaling parameter.  At each value of the scaling parameter the number of distinct features that exist may be counted for each homology group by simply counting the number of bars at that value.  For example, for the first simplicial complex there are twenty connected components and zero holes since the complex is starting with twenty vertices.  At the fourth simplicial complex there are three connected components and one hole.  The length of each bar indicates the length of each shape feature's existence.  \textbf{(b)}  Illustration of the persistence diagram.  The red and blue dots indicate show the persistent homology of the number of connected components and holes respectively.  For each dot, the $x$-coordinate indicates the value of the scaling parameter at which the shape feature appeared and the $y$-coordinate indicates the value of the scaling parameter at which the shape feature disappeared.  The distance from each point to the diagonal indicates the length of each shape feature's existence.  Since all features in $H_0$ existed at time zero, they are all above the point $x = 0$.  The features for $H_1$ came into existence at later times, so they are scattered at different points in the $x$-axis.} \label{summary-stats}
\end{figure}

Persistence diagrams are closely related to the summary statistic referred to as the barcode \cite{collins_barcode_2004}.  A barcode is simply a multiset of intervals $(\epsilon_a,\epsilon_b)$ in $\R^2$, with $\epsilon_a < \epsilon_b$, that keeps track of the appearance and disappearance of shape features across different scaling parameters.  Although the barcode is similar to that of the persistence diagram it does not require information about the diagonal.  Metrics and other properties of barcodes may be found in Carlsson et al. \cite{carlsson_persistence_2005}, and Zomorodian and Carlsson \cite{zomorodian_computing_2005}.  An example of how barcodes are constructed is found in Figure \ref{summary-stats}.


\subsection{Persistence Landscapes}  
\begin{figure}[htb!]
	\includegraphics[scale=0.30]{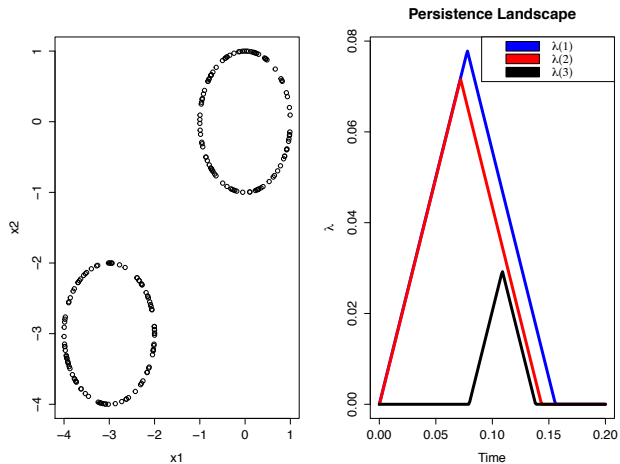}
	\caption{\textbf{Left:} A random sample of 200 points uniformly drawn from two different circles; 100 points were sampled from the circle centered at $(0,0)$ and 100 points were uniformly sampled from the circle centered $(-3,-3)$.  For these samples the Betti numbers for the connected components and holes are both two.  \textbf{Right:} The persistence landscape for the number of connected components.  The persistence landscapes are such that $\lambda_1 \geq \lambda_2 \geq \cdots \geq \lambda_n$.  Since the first two persistence landscapes are large, this gives evidence for the number of connected components being two, which is consistent with the example on the left.} \label{pl-img}
\end{figure}

Persistence landscapes are a new concept \cite{bubenik_statistical_2015} and are an alternative measure of persistent homology.  Persistence landscapes, while a statistic, in fact resides in a nicer mathematical space than persistence diagrams.  As such, they are more amenable to existing statistical methods (fully discussed in Section \ref{stat-pl}).  The interpretation of the persistence landscape is more subtle than the interpretation of either the persistence diagram and the barcode.  Rather than directly encoding information about the number of shape features in a homology group and their length of existence, the persistence landscape gives a measure of the number of features that simultaneously exist at a particular scaling parameter. An example of a persistence landscape is in Figure \ref{pl-img}.


\subsubsection{Mathematical properties of a persistence landscape}  A persistence landscape, $\Lambda$, is a sequence of piecewise continuous functions $\lambda_k : \R \to \R$.  Distance measures may be defined for persistence landscapes by integrating the functions, $\lambda_k$, and summing the result across all values of $k$.  Persistence landscapes can be treated as a random variable that takes values in a Banach space, which gives rise to the use of probabilistic results \cite{ledoux_probability_2011}.  Details and other results are covered in Bubenik \cite{bubenik_statistical_2015}.

\section{Statistical Inference with Topological Summary Statistics}

A significant amount of work has been performed for the purpose of extending statistical methods to Topological Data Analysis.  We are specifically interested in defining a clear notion of the mean and variance for persistent homology, developing testing methods that distinguish between the distributions from which topological summary statistics may have been sampled, and methods for determining which shape characteristics are statistically significant and which are topological noise.  Here, we review some of the statistical methods developed for both the persistence diagram and the persistence landscape.

It is generally assumed that the points of the sample $\mathcal{P} = \{X_1, \dots, X_n\} \subseteq \R^n$ are identically and independently distributed through some random process with distribution $\mathcal{F}_{X}$.  Applying the Topological Data Analysis workflow to $\mathcal{P}$ results in a topological summary statistic, $TS(\mathcal{P})$.  However, the distribution of our topological summary statistic may not follow the same distribution as the original sample.  Hence, in order to understand the distribution of the topological summary statistic it is necessary to understand the effect of the topological transformation on the distribution of the original data.  As a simple example of this effect, suppose $U$ is a random sample drawn from $\mbox{Uniform}(0,1)$ distribution, then $-\ln(U)$ is distributed as an $\mbox{exponential}(1)$ random variable.  In general, it is not clear what influence the topological transformation has on the distribution of the data.


\subsection{Fr\'{e}chet mean and variance of persistence diagrams} \label{mean-diagram}  The Fr\'{e}chet mean and variance of persistence diagrams are discussed in \cite{mileyko_probability_2011}.  Let $(\mathcal{D}_\mathcal{P}, \mathcal{B}(\mathcal{D}_\mathcal{P}), \mathcal{F}_{\mathcal{D}_\mathcal{P}})$ be a probability space on the space of persistence diagrams as defined in Section \ref{math-pd}, where $\mathcal{B}(\mathcal{D}_\mathcal{P})$ is the Borel $\sigma$-algebra on $\mathcal{D}_{\mathcal{P}}$, and $\mathcal{F}_{\mathcal{D}_\mathcal{P}}$ is a probability measure on this space.  In order to define the Fr\'{e}chet mean it is required that $\mathcal{F}_{\mathcal{D}_\mathcal{P}}$ have a finite second moment.  That is,
\begin{equation*}
	M_{\mathcal{D}_\mathcal{P}}(d) = \int_{\mathcal{D}_\mathcal{P}} W_p(d,e) \, d \mathcal{F}_{\mathcal{D}_\mathcal{P}}(e) < \infty \, ,
\end{equation*}
for a fixed diagram $d \in {\mathcal{D}_\mathcal{P}}$.  The Fr\'{e}chet variance is defined as
\begin{equation*}
	\mbox{Var}_{\mathcal{F}_{\mathcal{D}_\mathcal{P}}} := \inf_{d \in \mathcal{D}_\mathcal{P}} \Big\{ M_{\mathcal{D}_\mathcal{P}}(d) < \infty \Big\} \, ,
\end{equation*}
and the Fr\'{e}chet mean by
\begin{equation*}
	\mbox{E}_{\mathcal{F}_{\mathcal{D}_\mathcal{P}}} := \Big\{ d | M_{\mathcal{D}_\mathcal{P}}(d) = \mbox{Var}_{\mathcal{F}_{\mathcal{D}_\mathcal{P}}} \Big\} \, .
\end{equation*}
Since the definition of a Fr\'{e}chet mean is an infimum over a space, in general, the mean may not be unique.  To our knowledge, this is the only definition of a mean and variance for persistence landscapes.


\subsubsection{Algorithm for computing the Fr\'{e}chet mean and variance}  An algorithm for computing the Fr\'{e}chet mean is given in Turner et al. \cite{turner_frechet_2014} for the special case of  the $L^2$-Wasserstein metric and the distribution of the sample of persistence diagrams is a combination of Dirac masses \cite{dieudonne_treatise_1976}.  In this setting, the authors provide a law of large numbers, however they can only ensure that their algorithm converges to a local minimum.


\subsection{Hypothesis Testing for Persistence Diagrams}  A problem of particular interest in Topological Data Analysis is determining whether two subsets of $\R^n$ are the same.  Robinson and Turner  \cite{robinson_hypothesis_2013} present an argument that a necessary, but not sufficient, condition is that the underlying distribution of their persistence diagrams are the same.  To accomplish this, they develop a nonparametric permutation test to test for differences in the distributions of two different samples of persistence diagrams.  Rejecting a null hypothesis, $H_0$, that the two distributions are the same provides evidence that the two subsets, themselves, are different.  Details of the joint loss function that is employed, the reasons for developing a permutation test, and the justification for the output of their method being a $p$-value are in Robinson and Turner \cite{robinson_hypothesis_2013}.  


\subsection{Confidence Sets for Persistence Diagrams}  As with the majority of statistical applications, there is an interest in separating signal from noise.  As such, an interesting question in persistent homology is how to distinguish important shape features from topological noise.  The general working hypothesis in Topological Data Analysis is that features which exist  (i.e., persist) over large intervals of the scaling parameter are significant.  However, it is not always clear what constitutes a large interval, or if features that exist over a small interval are truly noise or of interest.  This first issue is addressed by Fasy et al. \cite{fasy_confidence_2014} who originated methods for computing a $1-\alpha$ confidence set for an estimated persistence diagram.  For instance, suppose $\mathcal{P}$ is a single sample taken from some distribution $\mathcal{F}$, then a persistence diagram $\hat{d}$ may be constructed from the data.  A value, $c_n$, is computed from the data so that when the vertical distance between a point and the diagonal is less than $\sqrt{2}c_n$ the lifespan of the feature is not different from zero.  

\subsection{Persistence Landscapes} \label{stat-pl} Persistence landscapes that are assumed to be random variables that take values in a Banach space are more amenable to the classical statistical theory of hypothesis testing and confidence intervals.  In Bubenik \cite{bubenik_statistical_2015} the vector space structure of the underlying $L^p$ space is used to construct a pointwise mean for the persistence landscape $\Lambda$.  That is, if we have samples $\mathcal{P}_1, \dots, \mathcal{P}_n$, with corresponding persistence landscapes $\Lambda^{1}, \dots, \Lambda^{n}$, then the mean landscape for the $k^{th}$ sequence is given by
\begin{equation*}
	\bar{\lambda}_k(t) = \frac{1}{n} \sum_{i = 1}^{n} \lambda^{i}_{k}(t) \, .
\end{equation*}
Using results from \cite{ledoux_probability_2011}, Bubenik \cite{bubenik_statistical_2015} shows that the Strong Law of Large Numbers holds for the mean persistence landscape and that they obey the Central Limit Theorem.  Chazal et al. \cite{chazal_stochastic} show that the convergence of the Central Limit Theorem is uniform.

\subsubsection{Hypothesis testing for persistence landscapes} An advantage of the persistence landscape is that it allows for hypothesis testing when samples are high-dimensional and non-linear.  In order to use persistence landscapes for hypothesis testing one has to use a functional - functions that map $\R^n$ to $\R$ - on the persistence landscapes.  When functionals satisfy certain conditions, Bubenik \cite{bubenik_statistical_2015} proved the Central Limit Theorem remains for the transformed persistence landscape.  Under this framework classical hypothesis testing procedures exist, and can be employed to distinguish between objects.  In other words, for a large number of samples, it is possible to use common $t$-tests or Hotelling's $T^2$ to distinguish between two subsets of $\R^n$.

\begin{figure}[tb!]
	\includegraphics[scale=0.75]{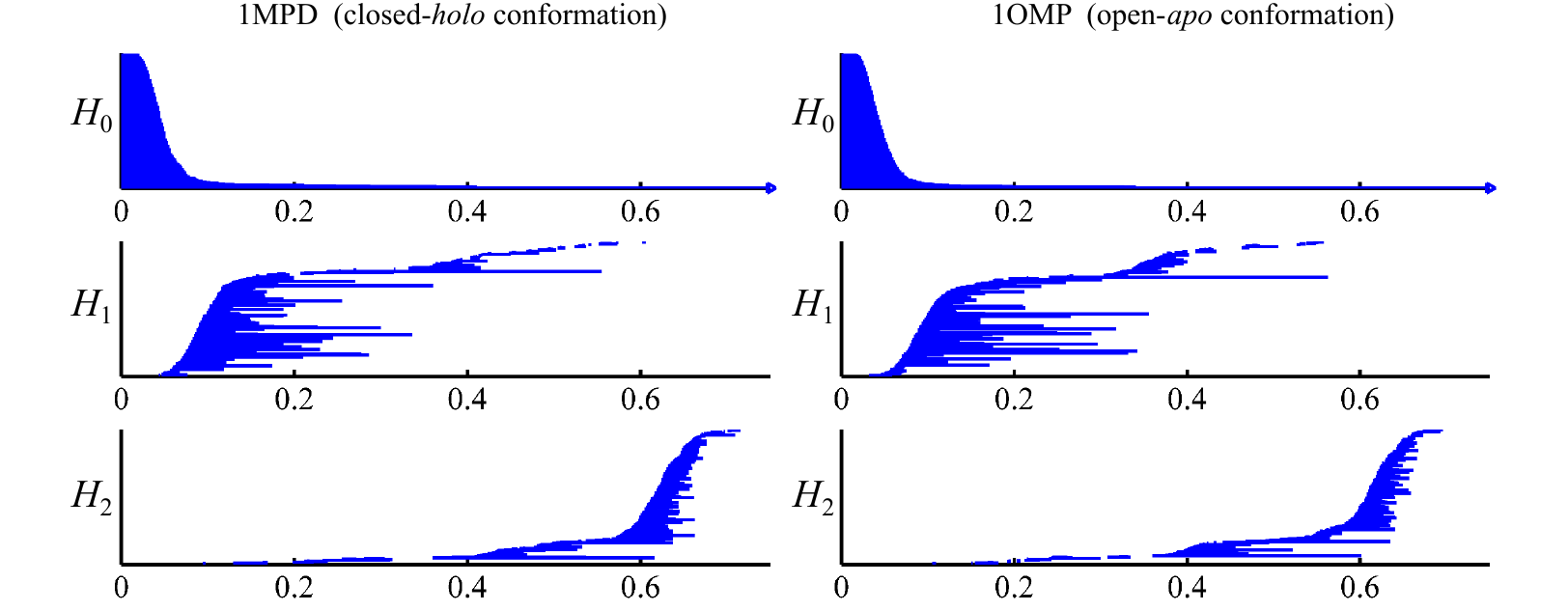}
	\caption{Barcodes for the first three homology groups of the closed conformation structure (left) and the open conformation structure (right) \cite{kovacev-nikolic_using_2014}. Slight differences between the two structures can be seen in the first homology group.  Differences between the other homology groups are more subtle.}
	\label{bc-mbp}
\end{figure}

It should be noted that applying a functional in fact, projects all of the information in a persistence landscape to a single point.  While this may result in a loss of information, this approach makes it easier to directly apply classical statistical tests, such as the $t$-test.

\section{Application of Topological Data Analysis for Studying the Conformation Space of the Maltose Binding Protein} \label{application}

\begin{figure}[tb!]
	\includegraphics[scale=0.75]{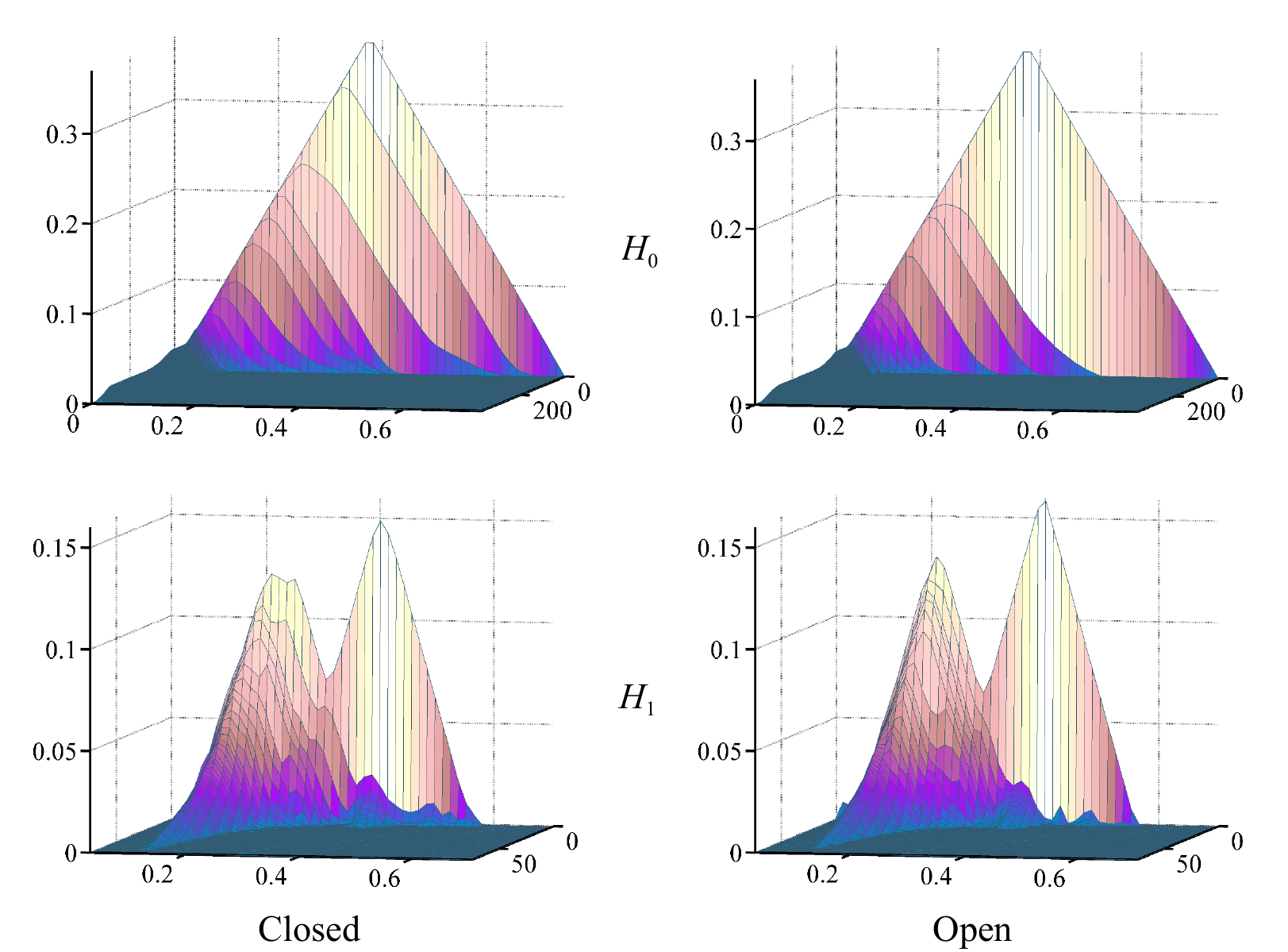}
	\caption{Persistence landscapes for the number of connected components, $H_0$ (top) and holes, $H_1$ (bottom) for the closed conformation structure (left) and the open conformation structure (right) \cite{kovacev-nikolic_using_2014}. (top) Observable differences between the closed and open persistence landscape are observed in the first homology group, $H_0$. The second persistence landscape of the closed structure attains a peak value of more than $0.30$ and attains that value at about $0.40$, while the second persistence landscape of the open structure attains a peak value at around $0.20$ and attains that value at about $0.30$.  For the first homology group, $H_1$, the initial peak of the first few persistence landscapes of the closed structure are uneven, while the same persistence landscapes in the open structure are smooth.}
	\label{pl-mbp}
\end{figure}

Technologies exist to capture proteins in a way that enables the study of their physical structure in three-dimensional space.  However, when a protein is captured its physical structure represents one of many possible shapes.  Many factors, including environmental or biological function, may influence the overall shape of a protein.  Kovacev-Nikolic et al. \cite{kovacev-nikolic_using_2014} consider the conformation space, or the space of possible shapes, of the maltose binding protein was studied using topological methods.  It is known that the maltose binding protein makes conformational changes when a ligand attaches.  If a protein is closed it is always prone to having a ligand attached.  One objective of Kovacev-Nikolic et al. \cite{kovacev-nikolic_using_2014} was to determine if statistically significant differences exist between the structures of the open and closed conformation space.

A sample of seven open structures and seven closed structures were obtained from the protein data bank \cite{protein_data_bank}.  The proteins were converted from their physical coordinates to a structure that considers the energy relationship between residues using the elastic network model.  Reasons for this conversion are discussed in \cite{kovacev-nikolic_using_2014}.  Persistent homology is then computed on each sample using the Vietoris-Rips complex, discussed in Section \ref{VRComplex}.  Figure \ref{bc-mbp} illustrates one of the computed barcodes for the first three homology groups, and Figure \ref{pl-mbp} illustrates the mean persistence landscapes for the first two homology groups.  In order to perform a hypothesis test, a functional is applied to each persistence landscape, resulting in a single value
\begin{equation} \label{pl-functional}
	X^{j}_{i,h} = \sum_{k = 0}^{\infty} \int_\R \lambda_k(t) \, dt \, ,
\end{equation}
where $1 \leq j \leq 7$, $i \in \{\mbox{open},\mbox{closed}\}$, and $0 \leq h \leq 2$ is the homology group.  This functional is simply the total area area under all of the persistence landscapes in the $k^{th}$ homology group.  Using these values, a permutation test is performed to compare the mean value of each homology group $H_0: \mu_{h,\mbox{open}} = \mu_{h,\mbox{closed}}$ against $H_a:  \mu_{h,\mbox{open}} \neq \mu_{h,\mbox{closed}}$.  The $p$-values for homology in both degree zero and one were computed to be $5.83 \times 10^{-4}$ giving evidence of a significant difference in the number of connected components and holes, while the $p$-value for the second homology group is 0.0396.  No corrections on the significance level were considered for multiple-testing, however the $p$-values for homology in degree zero and degree one are small enough to conclude differences in the means exist for these homology groups.

\section{Discussion}
Improvements in biotechnology have resulted in a massive growth of the amount and complexity of the data used in every field.  In this work we have introduced the untapped power of tools provided by Topological Data Analysis.  The TDA framework is well situated for studying datasets that are high-dimensional and complex.  Although there is evidence of success in applying these methods to both visualize high dimensional data and for classification, further extensions of statistical methods to Topological Data Analysis are needed.

\textcolor{white} \\

\textcolor{white} \\

\textcolor{white} \\

\textcolor{white} \\

\textcolor{white} \\

\textcolor{white} \\

\textcolor{white} \\

\textcolor{white} \\

\textcolor{white} \\

\textcolor{white} \\

\textcolor{white} \\

\textcolor{white} \\

\bibliographystyle{unsrt}
\bibliography{medina_kstate_2015} 

\begin{thebibliography}{10}

\bibitem{carlsson_topology_2009}
Gunnar Carlsson.
\newblock Topology and data.
\newblock {\em Bulletin of the American Mathematical Society}, 46(2):255--308,
  2009-04.

\bibitem{munkres_elements_1984}
James~R. Munkres.
\newblock {\em Elements of Algebraic Topology}.
\newblock Perseus, Reading, MA, 1984.

\bibitem{zomorodian_fast_2010}
Afra Zomorodian.
\newblock Fast construction of the vietoris-rips complex.
\newblock {\em Computer and Graphics}, page 263Ð271, 2010.

\bibitem{edelsbrunner_topological_2002}
{Letscher, D.} {Edelsbrunner, H.} and {Zomorodian, A.}
\newblock Topological persistence and simplification.
\newblock {\em Discrete \& Computational Geometry}, 28(4):511--533, 2002.

\bibitem{edelsbrunner_computational_2010}
Herbert Edelsbrunner and John Harer.
\newblock {\em Computational Topology: An Introduction}.
\newblock American Mathematical Society, 2010.

\bibitem{ghrist_barcodes:_2008}
Robert Ghrist.
\newblock Barcodes: The persistent topology of data.
\newblock {\em Bulletin of the American Mathematical Society}, 45(1):61--75,
  2008-01.

\bibitem{nanda_simplicial_2014}
Vidit Nanda and Sazadanovi\'{c} Radmila.
\newblock Simplicial models and topological inference in biological systems.
\newblock In Nata\v{s}a Jonoska and Masahico Saito, editors, {\em Discrete and
  Topological Models in Molecular Biology}, Natural Computing Series. Springer
  Berlin Heidelberg, 2014.

\bibitem{mileyko_probability_2011}
Yuriy Mileyko, Sayan Mukherjee, and John Harer.
\newblock Probability measures on the space of persistence diagrams.
\newblock {\em Inverse Problems}, 27(12):124007, 2011.

\bibitem{dudley_real_1989}
Richard~M. Dudley.
\newblock {\em Elements of Algebraic Topology}.
\newblock Chapman \& Hall, New York, NY, 1989.

\bibitem{collins_barcode_2004}
Anne Collins, Afra Zomorodian, Gunnar Carlsson, and Leonidas Guibas, J.
\newblock A barcode shape descriptor for curve point cloud data.
\newblock {\em Computer \& Graphics}, 28:881--894, 2004.

\bibitem{carlsson_persistence_2005}
{Gunnar} {Carlsson}, {Afra} {Zomorodian}, {Anne} {Collins}, and {Leonidas}~J.
  {Guibas}.
\newblock Persistence barcodes for shapes.
\newblock {\em International Journal of Shape Modeling}, 11(2):149--187, 2005.

\bibitem{zomorodian_computing_2005}
Afra Zomorodian and Gunnar Carlsson.
\newblock Computing persistent homology.
\newblock {\em Discrete Computational Geometry}, 33:249--274, 2005.

\bibitem{bubenik_statistical_2015}
Peter Bubenik.
\newblock Statistical topological data analysis using persistence landscapes.
\newblock {\em Journal of Machine Learning Research}, 16:77--102, 2015-01.

\bibitem{ledoux_probability_2011}
Michel Ledoux and Michel Talagrand.
\newblock {\em Probability in Banach Spaces}.
\newblock Classics in Mathematics. Springer-Verlag, 2011.

\bibitem{turner_frechet_2014}
Katharine Turner, Yuriy Mileyko, Sayan Mukherjee, and John Harer.
\newblock Fr{\'{e}}chet means for distributions of persistence diagrams.
\newblock {\em Discrete \& Computational Geometry}, 52(1):44--70, 2014-07.

\bibitem{dieudonne_treatise_1976}
Jean Dieudonne.
\newblock {\em Treatise on Analysis, Volume 2}.
\newblock Pure and Applied Mathematics (Book 10). Academic Press, 1976.

\bibitem{robinson_hypothesis_2013}
Andrew Robinson and Katharine Turner.
\newblock Hypothesis testing for topological data analysis.
\newblock {\em {arXiv} preprint {arXiv}:1310.7467}, 2013.

\bibitem{fasy_confidence_2014}
Brittany~T. Fasy, Fabrizio Lecci, Alessandro Rinaldo, Larry Wasserman,
  Sivaraman Balakrishnan, and Aarti Singh.
\newblock Confidence sets for persistence diagrams.
\newblock {\em The Annals of Statistics}, 42(6):2301--2339, 2014.

\bibitem{chazal_stochastic}
Frederic Chazal, Brittany~T. Fasy, Fabrizio Lecci, Alessandro Rinaldo, and
  Larry Wasserman.
\newblock Stochastic convergence of persistence landscapes and silhouettes.
\newblock {\em {arXiv}:1312.0308 [math.{ST}]}.

\bibitem{kovacev-nikolic_using_2014}
Violeta Kovacev-Nikolic, Peter Bubenik, Dragan Nikolic, and Giseon Heo.
\newblock Using cycles in high dimensional data to analyze protein binding.
\newblock {\em eprint {arXiv}:1412.1394}, page~21, 2014-12.

\bibitem{protein_data_bank}
F.~C. Bernstein, T.~F. Koetzle, G.~J. Williams, E.~F. Meyer, M.~D. Brice, J.~R.
  Rodgers, O.~Kennard, T.~Shimanouchi, and M.~Tasumi.
\newblock {The Protein Data Bank: a computer-based archival file for
  macromolecular structures.}
\newblock {\em Journal of molecular biology}, 112(3):535--542, May 1977.

\end{thebibliography}

\end{document}